\documentclass[10pt,letterpaper]{article}
\usepackage[top=0.85in,left=2.75in,footskip=0.75in]{geometry}

\usepackage{amsmath,amssymb}

\usepackage{changepage}

\usepackage[utf8x]{inputenc}

\usepackage{textcomp,marvosym}

\usepackage{cite}

\usepackage{nameref,hyperref}

\usepackage[right]{lineno}

\usepackage{microtype}
\DisableLigatures[f]{encoding = *, family = * }

\usepackage[table]{xcolor}

\usepackage{array}

\newcolumntype{+}{!{\vrule width 2pt}}

\newlength\savedwidth

\usepackage{tikz}
\usepackage[toc,page]{appendix}
\usepackage[colorinlistoftodos]{todonotes}
\usepackage{float}
\usepackage{subcaption}  
\usepackage[normalem]{ulem} 

\usepackage{algorithm}
\usepackage[noend]{algpseudocode}
\usepackage{authblk}
\usepackage{braket}

\newcommand{\liq}{LIQ$Ui|>$}

\raggedright
\usepackage[aboveskip=1pt,labelfont=bf,labelsep=period,justification=raggedright,singlelinecheck=off]{caption}

\makeatletter
\renewcommand{\@biblabel}[1]{\quad#1.}
\makeatother

\usepackage{lastpage,fancyhdr,graphicx}
\usepackage{epstopdf}
\fancyheadoffset[L]{2.25in}
\fancyfootoffset[L]{2.25in}
\begin{document}
\vspace*{0.2in}

\begin{flushleft}
{\Large
\textbf\newline{qTorch: The quantum tensor contraction handler} 
}
\newline
\\
E. Schuyler Fried\textsuperscript{1\dag},
Nicolas P. D. Sawaya\textsuperscript{1,2\dag},
Yudong Cao\textsuperscript{1},
Ian D. Kivlichan\textsuperscript{1},
Jhonathan Romero\textsuperscript{1},
Al\'{a}n Aspuru-Guzik\textsuperscript{1,3*}
\\
\bigskip
\textbf{1} Department of Chemistry and Chemical Biology, Harvard University, Cambridge, MA 02138, USA
\\
\textbf{2} Intel Labs, Intel Corporation, Santa Clara, CA 95054, USA
\\
\textbf{3} Canadian Institute for Advanced Research, Toronto, Ontario M5G 1Z8, Canada
\\
\bigskip

%
%
\dag These authors contributed equally to this work.





* alan@aspuru.com

\end{flushleft}
\section*{Abstract}

Classical simulation of quantum computation is necessary for studying the numerical behavior of quantum algorithms, as there does not yet exist a large viable quantum computer on which to perform numerical tests. Tensor network (TN) contraction is an algorithmic method that can efficiently simulate some quantum circuits, often greatly reducing the computational cost over methods that simulate the full Hilbert space. In this study we implement a tensor network contraction program for simulating quantum circuits using multi-core compute nodes. We show simulation results for the Max-Cut problem on 3- through 7-regular graphs using the quantum approximate optimization algorithm (QAOA), successfully simulating up to 100 qubits. We test two different methods for generating the ordering of tensor index contractions: one is based on the tree decomposition of the line graph, while the other generates ordering using a straight-forward stochastic scheme. Through studying instances of QAOA circuits, we show the expected result that as the treewidth of the quantum circuit's line graph decreases, TN contraction becomes significantly more efficient than simulating the whole Hilbert space. The results in this work suggest that tensor contraction methods are superior only when simulating Max-Cut/QAOA with graphs of regularities approximately five and below. Insight into this point of equal computational cost helps one determine which simulation method will be more efficient for a given quantum circuit. The stochastic contraction method outperforms the line graph based method only when the time to calculate a reasonable tree decomposition is prohibitively expensive. Finally, we release our software package, qTorch (Quantum TensOR Contraction Handler), intended for general quantum circuit simulation. For a nontrivial subset of these quantum circuits, 50 to 100 qubits can easily be simulated on a single compute node.

\section{Introduction}

Experimental hardware for quantum computing has been steadily improving in the past twenty years, indicating that a useful quantum computer that outperforms a classical computer may eventually be built. However, until a large-scale and viable quantum computer has been realized, numerically simulating quantum circuits on a classical computer will be necessary for predicting the behavior of quantum computers.

Such simulations can play an important role in the development of quantum computing by (1) numerically verifying the correctness and characterizing the performance of quantum algorithms \cite{Liquid,qhipster,qcmpi,qcwave,1012.6035}, (2) simulating error and decoherence due to the interaction between the quantum computer and its environment \cite{qhip-jctc,silva2008,geller2013,tomita2014}, and (3) improving our understanding of the boundary between classical and quantum computing in terms of computational power, for which recent efforts for characterizing the advantage of quantum computers over classical computers \cite{boixo18a,farhi16,boixo18b,chen18a,childs17,bouland18,pednault17,chen18b} serve as an example of this direction.


In this work, we consider the problem of quantum circuit simulation as one where we are given a quantum circuit and an initial state, with the goal of determining the probability of a given output state. Several approaches are possible for such simulation tasks. The most general method is to represent the state vector of an $N$-qubit state by a complex unit vector of dimension $2^N$ and apply the quantum gates by performing matrix-vector multiplications. This is essentially the approach adopted in, for example, \cite{Liquid,qcmpi,qhipster,projq,haner-steiger2017}. This method has the advantage that full information of the quantum state is represented at any point during the circuit propagation. However, the exponential cost of storing and updating the state vector renders it prohibitive for simulating circuits larger than $\sim$45 qubits. On the other hand, for a wide class of circuits with restricted gate sets and input states\cite{Gottesman_1998,quant-ph/0406196,Valiant:2001:QCS:380752.380785,quant-ph/0108010,2016PhRvL.116y0501B}, efficient classical simulation algorithms are available. For example, the numerical package Quipu \cite{6657072,garcia17} has been developed for taking advantage of prior results \cite{Gottesman_1998,quant-ph/0406196,2016PhRvL.116y0501B} on the stabilizer formalism to speed up general quantum circuit simulation. Finally, path integral-based methods \cite{gould-PI-2006} have also been proposed---though they do not improve the simulation cost, they lead to reduced memory storage requirements.


Other than considering the gate sets involved, an alternative perspective of viewing a quantum circuit is through its geometry or topology\cite{quant-ph/0403114,quant-ph/0511069}. Under this view, a quantum circuit is simulated via tensor network contractions. An advantage of viewing quantum circuits as tensor networks is that one can afford to ignore the particular kinds of quantum gates used in a circuit, and instead only focus on the graph theoretic properties. While general quantum circuits involving universal sets of elementary gates are likely hard to simulate on a classical computer \cite{mikeike11}, this geometric perspective sometimes allows for the efficient simulation of a quantum circuit with a universal gate set, provided that it satisfies certain graph theoretic properties. We note that at least one open source implementation of tensor network simulation for quantum circuits already exists, called TNQVM\cite{McCaskey_2016}, which can simulate tensor networks but also focuses on integrating algorithms with real quantum hardware. Aside from the field of quantum computation, tensor networks and related methods are an important and active area of research in the simulation of quantum mechanical problems in theoretical physics \cite{orus13,orus14,ran17}.

Among others, \emph{treewidth} is an important graph theoretic parameter that determines the efficiency of contracting a tensor network of quantum gates. A property of graphs that is actively studied in the graph theory literature \cite{Robertson1991153,Bodlaender2010259,BODLAENDER20111103,qbb}, the treewidth provides important structural information about a quantum circuit. Namely, if the circuit's underlying tensor network has treewidth $T$, it is shown in \cite{quant-ph/0511069} that the cost of simulating the circuit is $O(\text{exp}(T))$. In \cite{boixo18a} treewidth is also used for estimating the classical resource needed for simulating certain quantum circuits.

Motivated by the importance of tensor networks in quantum circuit simulation in general (and for example quantum computational supremacy tests in particular), it is useful to have a circuit simulation platform singularly dedicated to tensor network contraction. One immediate challenge in contracting tensor networks is to find an efficient contraction ordering, which relies on explicitly or implicitly finding a reasonable \textit{tree decomposition} of the underlying graph (definitions are further discussed in Section \ref{sec:prelim}). However, finding the optimal contraction ordering (or equivalently finding the minimum-size tree decomposition, or finding the treewidth of a graph) is NP-complete \cite{Arnborg:1987:CFE:37170.37183}. Therefore one must typically resort to heuristic methods when finding this decomposition \cite{amir10}.




For this study, we implemented a plug-and-play tensor network (TN) contraction code with two contraction schemes. Other schemes were attempted, but were significantly inferior to those that became part of the software package. However, there are likely other heuristic schemes that outperform our stochastic algorithm, and this is an avenue worth pursuing. For a large set of quantum circuits, our tensor network based methods are shown to be less costly than simulation of the full Hilbert space, by comparing to simulations using the \liq \space software package \cite{Liquid}. We emphasize again that the tests in this report give timing data for finding the expectation value of a measurement performed after implementing a quantum circuit, not for fully characterizing a circuit's final state.

The remainder of the paper is organized as the follows. Section \ref{sec:prelim} sets up the definitions and notations used in the paper. Section \ref{sec:contract} describes the heuristic methods used for contracting the quantum circuit tensor networks, along with other relevant details of the code implementation. Section \ref{sec:examples} presents the example quantum circuits used as benchmarks for demonstrating the performance of our contraction algorithms. Section \ref{sec:res} gives results of comparisons between the qTorch contraction methods, and between qTorch simulations and \liq's Hilbert space simulations.

\section{Preliminaries}\label{sec:prelim}

In this section, we provide an overview of relevant concepts and definitions. All graphs that we consider in this paper are undirected. We denote a graph as $G(V,E)$, consisting of the set of nodes $V=\{v_1, v_2,\cdots,v_n\}$ and edges $E\subseteq V\times V$.

Two relevant concepts are a graph's tree decomposition and treewidth \cite{Robertson1991153,amir10}. A \textit{tree decomposition} of a graph $G(V,E)$ is a pair $(S,T(I,F))$, where  $S=\{X_i|i\in I\}$ is a collection of subsets $X_i\subseteq V$ and $T$ is a tree (with edge set $F$ and node set $I$), such that $\cup_{i\in I}X_i=V$. Two nodes $X_i$ and $X_j$ are connected by an edge only if the intersection between $X_i$ and $X_j$ is not null. The \textit{width} of a tree decomposition  $(S,T)$ is $\max_{i\in I}|X_i|-1$. The \textit{treewidth} of a graph $G$ is the minimum width among all tree decompositions of $G$.

Another important concept in tensor network methods is the \textit{linegraph} of graph $G$, denoted by $L(G)$. $L(G)$ is itself an undirected graph, with each edge in $G$ corresponding to a node in $L(G)$. Two nodes in $L(G)$ are connected if and only if these two nodes' corresponding edges in $G$ are connected to the same node in $G$. There exists an optimal tree decomposition of $L(G)$ that provides the optimal contraction ordering of $G$ \cite{quant-ph/0511069}.

In the context of this work, a \textit{tensor} is defined as a data structure with rank $k$ and dimension $m$. More specifically, each tensor is a multidimensional array with $m^{k}$ complex entries. Each index may have a different dimension, though in this work each index has the same dimension $m=4$. A tensor $A_{i_{1},i_{2},i_{3},...i_{k}}$ has $k$ indices, which take values from $0$ to $m-1$.


A \textit{tensor contraction} is a generalized tensor-tensor multiplication. Here a rank $(x+y)$ dimension $m$ tensor $A$ and a rank $(y+z)$ dimension $m$ tensor $B$ are contracted into $C$, a rank $(x+z)$ dimension $m$ tensor. 
\begin{center}
\begin{equation}\label{eq:contr}
C_{i_{1},i_{2}...,i_{x},k_{1},k_{2}...,k_{z}} = \sum_{j_{1},j_{2}...,j_{y} \in \{0,..,m-1\}}{A_{i_{1},i_{2}...,i_{x},j_{1},...,j_{y}} B_{j_{1},j_{2}...,j_{y},k_{1},...,k_{z}}} 
\end{equation}  
\end{center}

Note that the number of floating point operations performed is $m^{x+y+z}$, exponential in the number of indices contracted on $y$ and the rank of the resulting tensor ($x+z$). It is important to point out that pairwise contractions are always optimal \cite{pfeifer14}. In other words, a function that contracts three or more nodes at a time will not achieve an improvement in scaling.

\begin{figure}[h]
\centering
\includegraphics[width=0.7\textwidth]{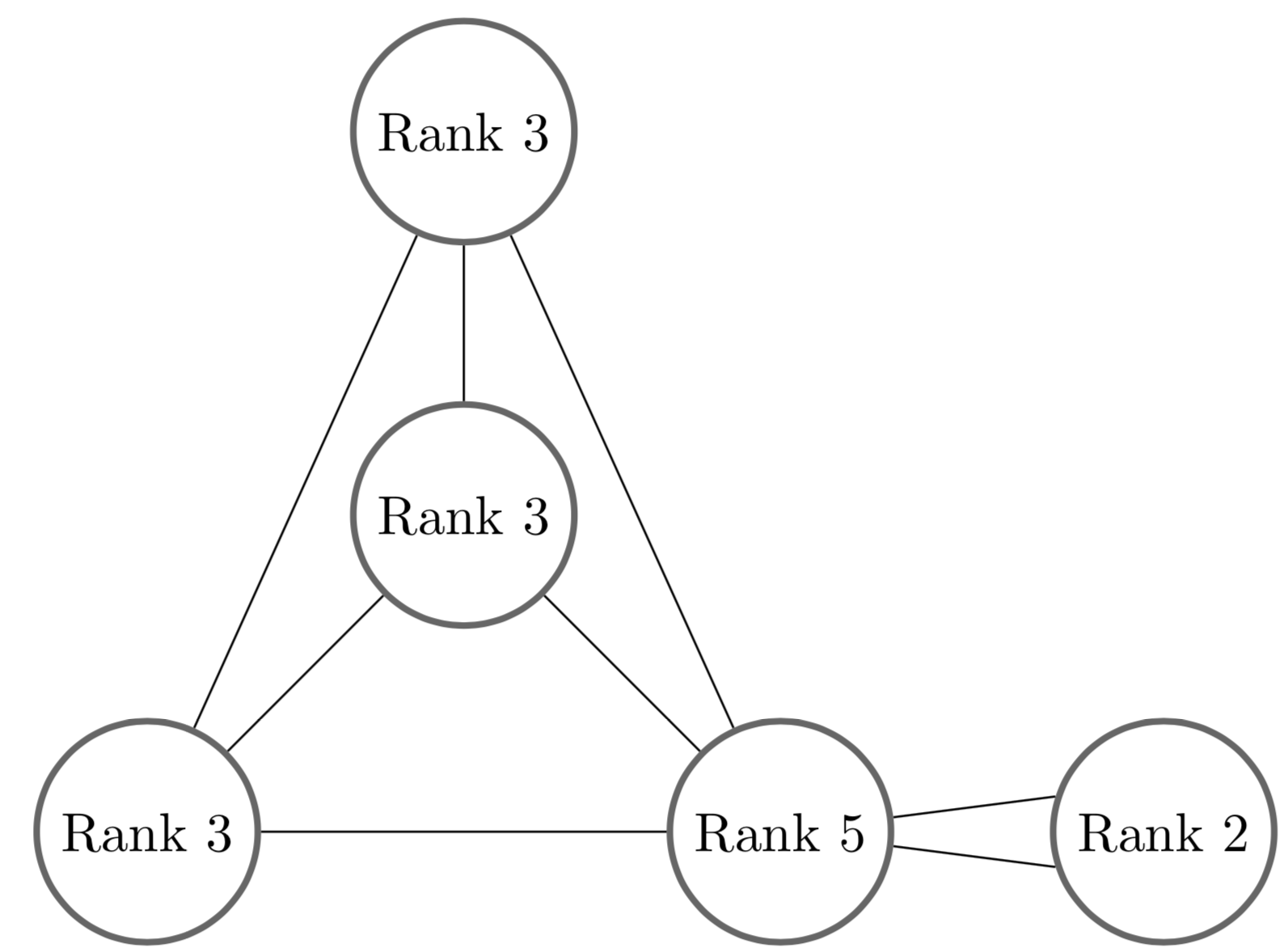}
\caption{An example of a tensor network. The number of edges (or wires) connecting to a tensor is equal to that tensor's rank. When an index (edge) is contracted by combining two tensors according to Equation \ref{eq:contr}, the two tensor are replaced by a new one. The number of scalar entries in the tensor scales exponentially in the number of edges to which it connects. In general it is not trivial to choose an efficient contraction ordering that minimizes the total number of floating point operations.}
\label{fig:tn}
\end{figure}

A \textit{tensor network} is a graph $G = (V,E)$ with tensors as vertices, and edges labeled by a set of indices. The rank of each tensor is given by the number of edges connected to it. An edge from one tensor to another indicates a contraction between the two tensors, and multiple connected edges indicate a contraction on multiple indices. Fig \ref{fig:tn} shows an example of a tensor network. Note that a tensor may have open edges, \textit{i.e.} edges that do not connect to any other tensor, though this possibility is not allowed in the current version of qTorch.

 

A contraction ordering or contraction scheme determines the order in which the tensor network is contracted. The ordering chosen for the contraction will greatly affect the computation and memory requirements, because some contraction orderings can result in much larger intermediate tensors than others. Although in this work we focus on contracting the tensor network to a scalar that equals the expectation value of the quantum circuit's measurement, the goal of a tensor network algorithm is often \textit{not} to contract it to a scalar\cite{orus13,ran17}. An example of this is the infinite tensor networks used to study periodic systems in condensed matter physics.

\begin{figure}[h]
\centering
\includegraphics[width=0.3\textwidth]{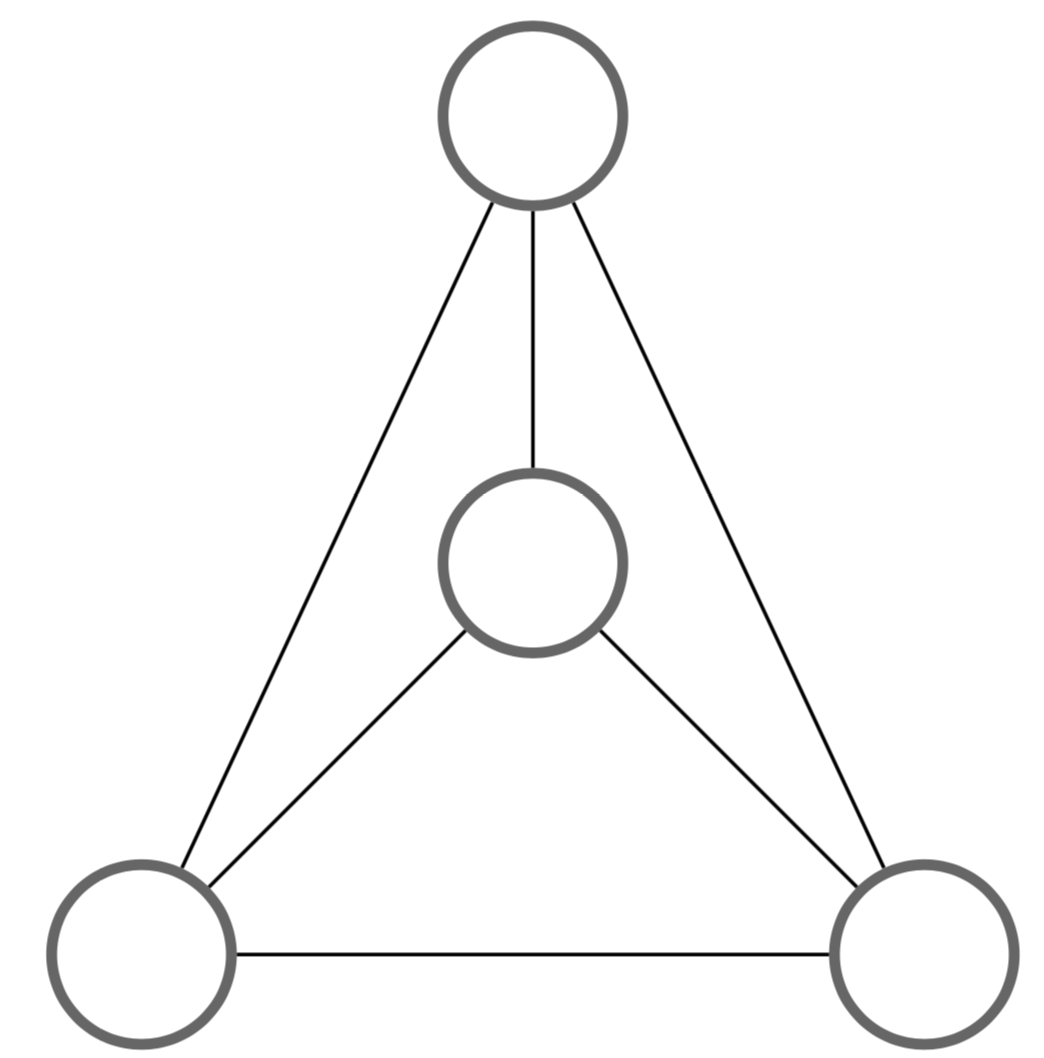}
\caption{A tetrahedral graph illustrates why it is often unavoidable to form tensors of higher rank while contracting a tensor network. In this example, contracting any of the six edges produces a tensor of rank 4, even though all of the original tensors were of rank 3.}
\label{fig:tetr}
\end{figure}

An important goal is to avoid tensors of large intermediate rank when contracting the network, as floating point operations grow exponentially with tensor rank. However, it is often the case that increasing the tensor rank is unavoidable. A simple example of this issue is a tetrahedral graph of rank 3 tensors (Fig \ref{fig:tetr}), which cannot be contracted without forming intermediate tensors of rank greater than 3. The larger the treewidth of $L(G)$ is, the more one will be forced to create new tensors of higher rank as the network is contracted, greatly increasing the computational cost \cite{quant-ph/0511069}. 

 

We note that tensor network methods are commonly used to efficiently find approximate solutions---indeed this is often the main source of a TN's utility. In approximate TN methods, the space of the smaller values of the eigenspectrum are removed, after e.g. performing a singular value decomposition on the tensor\cite{orus13,ran17}. This allows one to contract to form a larger tensor, then renormalize its size before continuing to contract the network. Though this strategy is useful in the context of quantum circuits as well, the current version of qTorch is limited to numerically exact contractions of the tensor network.

Before contracting, the tensor network graph must first be created from a quantum circuit, a procedure that has been summarized in previous work \cite{quant-ph/0511069}. Each node on the graph represents one of the following: An initial state of the qubit (usually $|0 \rangle \langle 0|$), a gate operation, or a measurement. The initial density matrix is represented as a rank 1 dimension 4 tensor (i.e. a vector), $[\rho_{|0 \rangle \langle 0|};\rho_{|0 \rangle \langle 1|};\rho_{|1 \rangle \langle 0|};\rho_{|1 \rangle \langle 1|}]$. Measurement nodes are rank 1 as well. All indices in the graph are dimension 4, regardless of rank. Nodes corresponding to quantum gates are represented in superoperator form. Hence a gate $U$ which acts on the quantum state as $\rho \rightarrow U \rho U^\dag$ is represented by the superoperator $\tilde{U}$. The same operation can be expressed as $\tilde{\rho} \rightarrow \tilde{U}\tilde{\rho}$, where $\tilde{\rho}$ is the Lindblad representation of the density operator. Single qubit gates correspond to rank 2 tensors and two-qubit gates correspond to rank 4 tensors. The graph's connectivity is identical to the connectivity of the original quantum circuit.

We end this section with explicit examples of tensors for standard quantum circuit components. Tensors for other circuit components can be viewed in the source code for qTorch.

The initial state $|0 \rangle\langle 0|$ corresponds to the tensor

\begin{equation}
\tilde{\rho}_0 = [1,0,0,0].
\end{equation}

Superoperator tensors for the Pauli matrices are

\begin{equation}
\tilde{X} =
\begin{bmatrix}
   0 & 0 & 0 & 1 \\
   0 & 0 & 1 & 0 \\
   0 & 1 & 0 & 0 \\
   1 & 0 & 0 & 0 \\
\end{bmatrix};
\tilde{Y} = 
\begin{bmatrix}
   0 & 0 & 0 & 1 \\
   0 & 0 & -1 & 0 \\
   0 & -1 & 0 & 0 \\
   1 & 0 & 0 & 0 \\
\end{bmatrix};
\tilde{Z} = 
\begin{bmatrix}
   1 & 0 & 0 & 0 \\
   0 & 1 & 0 & 0 \\
   0 & 0 & 1 & 0 \\
   0 & 0 & 0 & 1 \\
\end{bmatrix}.
\end{equation}

The CNOT gate is represented as a sparse rank 4 tensor for which only the following entries are nonzero:

\begin{align}
\tilde{U}_{CNOT}:
\tilde{U}_{0000} =
\tilde{U}_{0101} =
\tilde{U}_{0202} =
\tilde{U}_{0303} =
\tilde{U}_{1011} =
\tilde{U}_{1110} =
\tilde{U}_{1213} =
\tilde{U}_{1312} = \nonumber \\
\tilde{U}_{2022} =
\tilde{U}_{2123} = 
\tilde{U}_{2220} =
\tilde{U}_{2321} =
\tilde{U}_{3033} =
\tilde{U}_{3132} =
\tilde{U}_{3231} =
\tilde{U}_{3330} = 1
\end{align}

Finally, the nodes for measurement are rank 1 tensors. $\tilde{M}_X$, $\tilde{M}_Y$, and $\tilde{M}_Z$ correspond respectively to determining expectation values for measurements in the $X$, $Y$, and $Z$ bases. Note that using $\tilde{M}_X$, $\tilde{M}_Y$, or $\tilde{M}_Z$ is equivalent to inserting a Pauli gate at the end of the circuit before implementing $\tilde{M}_{Trace}$.
\begin{equation}
\tilde{M}_{Trace} =
\begin{bmatrix}
1 \\ 0 \\ 0 \\ 1
\end{bmatrix};
\tilde{M}_{X} =
\begin{bmatrix}
0 \\ 1 \\ 1 \\ 0
\end{bmatrix};
\tilde{M}_{Y} =
\begin{bmatrix}
0 \\ i \\ -i \\ 0
\end{bmatrix};
\tilde{M}_{Z} =
\begin{bmatrix}
1 \\ 0 \\ 0 \\ -1
\end{bmatrix}
\end{equation}

\section{Contraction schemes and implementation details}\label{sec:contract}
For many problems in quantum physics to which matrix product states (MPS) or other tensor network methods have been applied, an efficient contraction scheme is often obvious from the underlying structure of the Hamiltonian \cite{orus14}. However, efficient contraction schemes are not available for arbitrary tensor networks. A general heuristic contraction scheme is important for the simulation of general quantum circuits, when one does not know a priori the topological properties of the underlying tensor network problem. 

\subsection{Contraction schemes}

qTorch implements two algorithms for determining the contraction ordering. For what we call the line graph (\textit{LG}) method, outlined in Algorithm 1, we first create the line graph of the quantum circuit's graph. Then, the software package QuickBB \cite{qbb} is used to determine an approximately optimal tree decomposition of this linegraph. The resulting tree decomposition is used to define the order of contraction. This linegraph-based approach was first described by Markov and Shi \cite{quant-ph/0511069}.

QuickBB uses a so-called anytime algorithm, meaning that it can be run for an arbitrary amount of time, such that when the program is stopped it provides the best solution found thus far. The algorithm is based on the branch and bound (B\&B) method, though the authors use several techniques based on modern graph theory to improve efficiency in the pruning and propagation steps, making QuickBB faster at finding low-width tree decompositions than vanilla B\&B.

The second contraction scheme is a simple stochastic procedure we refer to as \textit{Stoch} (Algorithm 2). First, a wire is randomly chosen. If the rank of the contracted tensor is higher than the highest rank of the two nodes, plus a given threshold, the contraction is rejected. After a fixed number of rejected contraction attempts, the threshold is relaxed.

\begin{algorithm}
\caption{Contraction via TD of L(G)}\label{td}
\begin{algorithmic}[1]

\State Create line graph L(G) of graph G
\State $\pi \gets$ (Calculate approx. optimal elimination ordering of L(G))
\State Eliminate wires of G in order $\pi$

\end{algorithmic}
\end{algorithm}

\begin{algorithm}
\caption{Simple stochastic contraction}\label{simple}
\begin{algorithmic}[1]
\State Define $G \gets$ The tensor network Graph
\State $Threshold \gets -1$
\State Define $MaxRejections$ $\gets$ Maximum Number of Rejections

\Repeat
  \State Choose a random wire $w$
  \State ($A,B$) $\gets$ (Nodes of $w$)
  \State Cost $\gets$ rank($C$) $-$ max(rank($A$),rank($B$))

  \If{Cost $\leq$ Threshold}
      \State Contract $w$ to form node $C$
      \State $Rejections$ $\gets$ 0
      \State $Threshold$ $\gets$ -1
      \State Update $G$
  \Else
      \State $Rejections \gets Rejections + 1$
      \If{$Rejections > MaxRejections$}
      	\State $Threshold \gets Threshold + 1$
        \State $Rejections \gets 0$
        \State Continue
      \EndIf
  \EndIf
\Until{Graph completely contracted}

\end{algorithmic}
\end{algorithm}


\subsection{Threading}
The tensor contractions are parallelized using the C++ standard library's \texttt{std::thread} class. A particular tensor-tensor contraction is parallelized if the cost of contracting a pair is higher than a provided threshold. We implement other parallelization schemes at the network level, i.e. splitting up the nodes into separate groups to compute on different threads, but the vast majority of the parallelization speedup comes from threading the tensor--tensor contractions. Currently, qTorch does not support parallelization across multiple compute nodes within a cluster, but it allows the user to specify the number of threads (default of 8).

\subsection{Estimating the answer string}\label{ssec:ansstr}


qTorch computes expectation values of the form $\langle \psi | M | \psi \rangle$, where $M$ is a measurement operator such as a Pauli string, and $|\psi\rangle$ is the quantum state produced by the circuit. If instead one wishes to capture all the information of this final state of $n$ qubits, it generally requires $O(2^n)$ repetitions of the algorithm. However, many quantities of interest may be calculated efficiently. For instance, the probability that one measurement operator (e.g. a Pauli string) will provide a particular outcome can be estimated in just one contraction of the tensor network, a result essential to simulating the variational quantum eigensolver (VQE) \cite{Peruzzo.NC.5.4213.2014,weckervqe15,Yung.SR.4.3589.2014,Mcclean.NJP.18.023023.2016}.

qTorch provides a heuristic scheme to output a high-probability answer string from the circuit, which we summarize here. Though this scheme is not used for the results presented in Section \ref{sec:res}, it may be useful in the future for simulating algorithms (like QAOA) where the goal is to estimate a most likely bit string.

The scheme is implemented as follows. First we run one simulation, and measure in the computational basis to project the first qubit into 0 or 1. Based on the resulting expectation value from the simulation, we choose the value for the first qubit that has the greater probability. If the 0 and 1 are equally likely, one is chosen randomly. In the next full contraction, we set the resulting binary value as the measurement outcome for the first qubit in the next simulation, and repeat with a projective measurement on the second qubit. We continue this process for the rest of the qubits. As we show below, this method often gives a good approximation of the most likely final computational basis state. In original tests on 3-regular graphs of 30 vertices, the scheme (used on Quantum Approximate Optimization Algorithm [QAOA] circuits) gave bit strings that provided good estimates to the solution of the Max-Cut problem (average approximation ratio of 94\% compared to the exact brute force solution). 


As a way to test the general applicability of this scheme, we performed some tests on more general circuits than the QAOA problem. These tests are meant to provide some early insight into how useful this heuristic would be for estimating the most likely bit string of a quantum algorithm, for the users who are interested in running this string estimation subroutine. We note that it is abundantly clear that in many cases this scheme does not produce a string closest to the most likely string---indeed, if it was a generally accurate scheme then we would have no need for a quantum computer.

In the remainder of this section, we consider the most likely bit string of the final state $|\psi\rangle=\sum_i\psi_i|i\rangle$, which we define as $\text{argmax}_i|\psi_i|^2$, where the vectors $\{|i\rangle\}$ are in the computational basis. We apply a unitary of the form
\begin{equation} \label{eq:unitary}
U_p({\boldsymbol\beta},{\boldsymbol\gamma})=
\prod_{j=1}^p
\text{exp}\left(i\beta_j\sum_{i=1}^n X_i\right)
\text{exp}\left(i\gamma_j{D}_j\right)
\end{equation}
where the matrix ${D}_j$ is a diagonal matrix with entries chosen randomly from the integers $\{1,\cdots,nm\}$. Here $m$ is a parameter that could be interpreted as the number of clauses, if this were a QAOA problem. The elements of the $p$-dimensional vectors $\boldsymbol\beta$ and $\boldsymbol\gamma$ are drawn uniformly from $[0,\pi]$ and $[0,2\pi]$ respectively. We use the construction of $U_p({\boldsymbol\beta},{\boldsymbol\gamma})$ to emulate the form of parametrized unitary operations used in QAOA with the same $p$. Starting from the uniform superposition over all $2^n$ bit strings $|s\rangle$, we apply $U_p$ to compute the final state $|\Psi\rangle=U_p|s\rangle=\sum_{i=0}^{2^n-1}\psi_i|i\rangle$. Let $p_i=|\psi_i|^2$ denote the probability distribution associated to the QAOA-like output state $|\Psi\rangle$. We ran 10,000 trials (with n=6, m=10, and p=2) using equation \ref{eq:unitary}, and ranked the result by how many bit strings in the true state were more likely than our outputted bit string. 

Conceptually, our likely string estimation algorithm can be thought of as falsely assuming that the output state is a product state. Suppose we apply our algorithm to the state $|\Psi\rangle$. The product state then reads
\begin{equation}\label{eq:psi_}
|\Psi'\rangle=(\alpha_1|0\rangle+\beta_1|1\rangle)\otimes
(\alpha_2|0\rangle+\beta_2|1\rangle)\otimes\cdots
(\alpha_n|0\rangle+\beta_n|1\rangle)
\end{equation}
where $|\alpha_k|^2$ is the probability of $|0\rangle$ that the algorithm obtains at the $k^\text{th}$ step, with an analogous definition for $\beta_k$. With this conceptual framing, we also numerically study the 1-norm distance $\|{\bf p}'-{\bf p}\|_1$ between the approximate distribution ${\bf p}'$ which the algorithm effectively assumes and the actual distribution ${\bf p}$.

\begin{figure}[H]
	\centering
 	\includegraphics[width=1.0\textwidth]{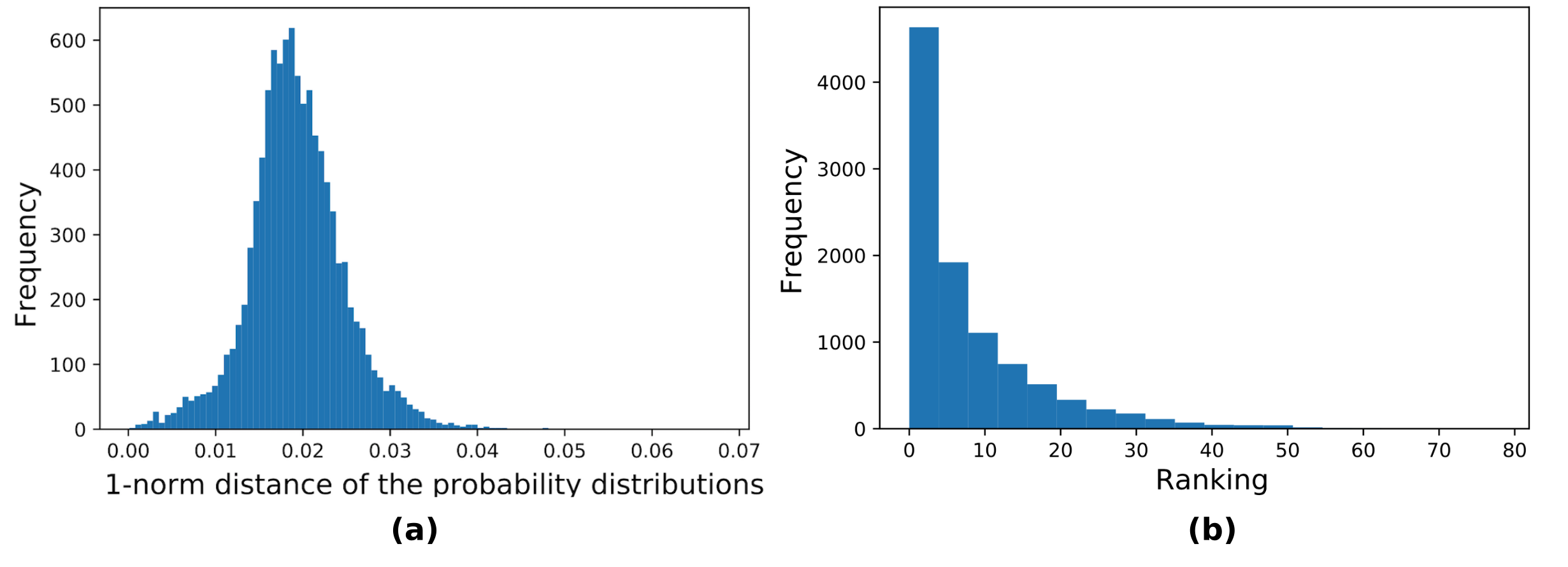}
\caption{Results from implementing 10,000 trials of Equation \ref{eq:unitary}. We use $n=6$ qubits, $m=10$, and $p=2$. \textbf{(a)} The histogram plots how close the method's output string is to the actual most likely string. qTorch's procedure for the ``estimate'' is given in the text. The horizontal axis \emph{Ranking} is the number of computational basis states in $|\Psi\rangle$ with higher probability than the estimated string---a lower number ranking indicates a better estimate. \textbf{(b)} Distribution of the 1-norm distance between the approximate distribution ${\bf p}'$ arising from the product state approximation $|\Psi'\rangle$ in and the distribution  ${\bf p}$ arising from the exact state $|\Psi\rangle$. }
\label{fig:prodstate}
\end{figure}

The results are shown in Fig \ref{fig:prodstate}. Here we use the number of qubits $n=6$, with parameters $m=10$ and $p=2$. Fig \ref{fig:prodstate} (a) shows that most of the time our algorithm produces a high ranking bit string---roughly 90\% of the time the output of the algorithm is among the top 10\% most likely bit strings. Fig \ref{fig:prodstate} (b) shows that the 1-norm distance between the approximate and exact distributions is less than 0.1 for nearly all of the data points. These results suggest that our heuristic for estimating an output bit string will produce acceptable estimates for some circumstances---in QAOA for instance, where one might be interested in a good approximate (as opposed to exact) solution to the constraint satisfaction problem. 

\subsection{A note on noise}
It is possible to simulate noise within the quantum circuit model, by mapping a noise model onto a set of one-qubit or multi-qubit operators \cite{mikeike11,kraus83}.

Note that any quantum operation can be expressed in terms of Kraus operators $\{E_j\}$
\begin{equation}\label{eq:kraus}
\rho \rightarrow \sum_j E_j \rho E_j^\dag
\end{equation}

where $\{E_j\}$ are called Kraus operators \cite{kraus83}, and $\sum_j E_j E_j^\dag=1$ because for our purposes the noise process is assumed to be trace-preserving. A noise model can be expressed in terms of such Kraus operators, which can in turn be expressed as superoperators for insertion into the quantum circuit's tensor network.

The most commonly used approximations assume that noise on different qubits is uncorrelated, which allows for single-qubit ``noise gates'' to be used. In this case, because rank 2 tensors can always be contracted without increasing the rank of the resulting tensors, the cost of simulating the resulting ``noisy'' quantum circuit would not substantially increase. One common and easily implementable approximation is the Pauli twirl approximation, which approximates a noise process purely in terms of Pauli rotations \cite{silva2008,geller2013,tomita2014,qhip-jctc}, and therefore can be implemented with the built-in quantum gates of qTorch.

A more physically realistic noise model would assume correlated noise \cite{aharonov06,preskill13}, which necessitates the insertion of noise gates that operate on at least two qubits. In this case, the tree width of the circuit's underlying line graph, and hence the complexity of the problem, would increase in all but the most trivial cases.

qTorch does not incorporate built-in noise gates. Instead, we include functionality that allows for user-defined gates.

\section{Circuit simulations} \label{sec:examples}

Here we describe the classes of quantum circuits that were simulated for this work.

\subsection{QAOA / Max-Cut}\label{ssec:main-qaoa}

The quantum approximate optimization algorithm (QAOA) was recently developed \cite{farhi14a}, for the purpose of demonstrating quantum speedup for combinatorial problems on low-depth quantum circuits. Given a constraint satisfaction problem (CSP), a QAOA quantum circuit produces an output that provides approximate solutions. Several aspects of QAOA have been studied since its introduction, including its application to different classes of CSP, implementations of different classical optimization routines, and numerical and analytical comparisons to classical algorithms \cite{farhi14a,farhi14b,farhi16,wecker16,lin16,gg17}.


We use qTorch to simulate QAOA for the Max-Cut problem, a combinatorial problem that has been a focus of QAOA \cite{farhi14a}. Given an arbitrary undirected graph, the goal of Max-Cut is to assign one of two colors to each node, so as to maximize the number of cuts. A cut is any edge that connects two nodes of different color.

In QAOA, a set of constraints is mapped to a an objective function represented by a set of operators. Specifically for the Max-Cut problem, the object function is

\begin{equation} \label{eq:maxcutobj}
C = \sum_{\langle ij\rangle }{C_{\langle ij \rangle}},
\end{equation}    

with 

\begin{equation} \label{eq:pairwise}
C_{\langle ij\rangle } = \frac{1}{2}(1-\sigma^{z}_{i}\sigma^{z}_{j}),
\end{equation}
where $\langle ij\rangle$ represents the edge between nodes $i$ and $j$, $\sigma_k^z$ is the Pauli-Z operator on qubit $k$, and each node in the underlying Max-Cut graph (which is related to but not the same as the quantum circuit's graph) corresponds to one qubit in the quantum circuit.

Define two operators $U(C,\gamma)$ and $U(B, \beta)$ as
\begin{equation}
U(C,\gamma) = e^{-i\gamma C} = \prod_{m=1}^{n}{e^{-i\gamma C_{m}}}
\end{equation} 

and 

\begin{equation}
U(B, \beta) = e^{-i\beta B} = \prod_{k=1}^{q}{e^{-i\beta \sigma_k^x}}.
\end{equation}  

where $B=\sum_{k=1}^q \sigma_k^x$, $\sigma_x$ is the Pauli-X operator, $q$ is the number of qubits, and $n$ is the number of clauses (for Max-Cut this is the number of edges). These two operators are applied $p$ times (with different $\gamma$ and $\beta$ allowed for each step), with a larger $p$ providing a better approximation. The $\gamma$ and $\beta$ parameters are modified with a classical optimization routine to maximize the cost function. The cost function is evaluated after each measurement, with the bit string that resulted from the measurement.

To generate the graphs for the underlying Max-Cut problem, we made random $k$-regular graphs by placing edges randomly throughout a given vertex set to satisfy a given regularity, before ensuring that disconnected graphs are rejected. QAOA/Max-Cut Quantum circuits based on these graphs are then constructed.

In the numerical results of this paper, we report only the timing for a single contraction of each quantum circuit. A full analysis of QAOA is beyond the scope of this work. However, we note that once the graphs have been created, qTorch currently has the functionality to optimize the QAOA angles using the classical optimization library NLopt \cite{nlopt}. Finally, one can use qTorch to estimate a Max-Cut for the randomly-generated graph, using the most-likely bit string estimation method described above.

\subsection{Hubbard Model}

Quantum simulation of fermionic systems is one of the most relevant applications of quantum computers, with direct impact on chemistry and materials science, including for the design of new drugs and materials. Among all the algorithms proposed for quantum simulation of fermions, the quantum variational algorithm (VQE) and related approaches \cite{Peruzzo.NC.5.4213.2014,weckervqe15,Yung.SR.4.3589.2014,Mcclean.NJP.18.023023.2016} are arguably the most promising for near-term hardware because they have lower circuit depth requirements \cite{mcclean2016hybrid,Omalley.PRX.6.031007.2016}. We note that many types of chemistry-related circuits can be prepared with the software package OpenFermion \cite{openfermion}.

In the VQE algorithm, a quantum computer is employed to prepare and measure the energy of quantum states associated with a parameterized quantum circuit. The approximate ground state of a Hamiltonian is obtained by variationally minimizing the cost function (corresponding to e.g. the molecular energy) with respect to the circuit parameters using a classical optimization routine. This hybrid quantum-classical approach offers a good compromise between classical and quantum resources. Classical simulations of the VQE algorithm for tens of qubits could provide insights into the complexity of the circuits used for state preparation and help design better ansatzes for the quantum simulation of fermions.

As an example of a VQE simulation, we used qTorch to classically simulate variational circuits employed for the quantum simulation of 1D Hubbard lattices. We consider half-filled Hubbard models on $N$ sites, with periodic boundary conditions.


To construct variational circuits for these systems, we considered the variational ansatz introduced by Wecker et al \cite{weckervqe15}. In this case, the Hubbard Hamiltonian is divided as $H=h_h+h_U$, where $h_h$ is the sum of hopping terms in the horizontal dimension and $h_U$ is the repulsion term. The variational circuit is constructed as a sequence of unitary rotations by terms in the Hamiltonian with different variational parameters, with the sequence being repeated $S$ times. In each step, there are two variational parameters, $\theta^{b}_{U}$ and $\theta^{b}_{h}$, where $b=1,\cdots,N$ such that

\begin{equation}\label{variationalC}
U = \prod^{S}_{b=1} \left[ U_U \left(\frac{\sigma^{b}_U}{2}\right) U_h(\theta^{b}_{h})  U_U\left(\frac{\theta^{b}_U}{2}\right) \right]
\end{equation}

where $U_X(\theta_X)$ denotes a Trotter approximation to $\exp(i\theta_X h_X)$, and $X$ can be $U$ or $h$. For our numerical simulations, we employed the variational circuit of Eq. \ref{variationalC} with $S=1$ using a 1-step Trotter formula for all the $U_X(\theta_X)$ terms. Notice that this is approximate for the $h_h$ term, which comprises a sum of non-commuting terms. We also assigned the value of 1 to all variational amplitudes. The corresponding unitary was mapped to a quantum circuit using the Jordan-Wigner transformation and the circuit was generated using a decomposition into CNOT gates and single-qubit rotations \cite{mikeike11,Whitfield.MP.109.735.2011}. 

\section{Results}\label{sec:res}

Simulations were performed on the NERSC Cori supercomputer, using one "Knights Landing" (KNL) node per simulation, each of which contains 68 cores and 96 GB of memory. Each \liq \space simulation was run on a full node as well, using Docker \cite{docker2014}.
The free version of \liq \space allows for the simulation of 24 qubits. Because full Hilbert space simulation scales exponentially regardless of the quantum algorithm's complexity, we would not have been able to simulate more than $\sim$31 qubits on one of these compute nodes. For each set of parameters (regularity and number of vertices/qubits) 50 instances of Max-Cut/QAOA circuit were created. For higher qubit counts and higher regularities, only a subset of these circuits were completed, since many simulations exceeded memory capacity. In this section, \textit{LG} or \textit{qTorch-LG} refer to the use of qTorch with the linegraph-based contraction, \textit{Stoch} or \textit{qTorch-Stoch} refer to qTorch with stochastic contraction. To determine a qTorch-LG contraction ordering, QuickBB simulations were run for an arbitrary time of 3000 seconds for each quantum circuit. The plotted qTorch results do not include the QuickBB run time.

We note that \liq \space implements a thorough set of important optimizations, which makes it a fair benchmark against which to compare qTorch. For example, \liq \space fuses many gates together before acting on the state vector, and uses sparse operations. qTorch, on the other hand, does not yet use sparsity at all (even when the circuit consists primarily of sparse CNOT gates), which is one of several optimizations that we expect would further improve performance.

\begin{figure}[H]
	\centering
 	\includegraphics[width=0.8\textwidth]{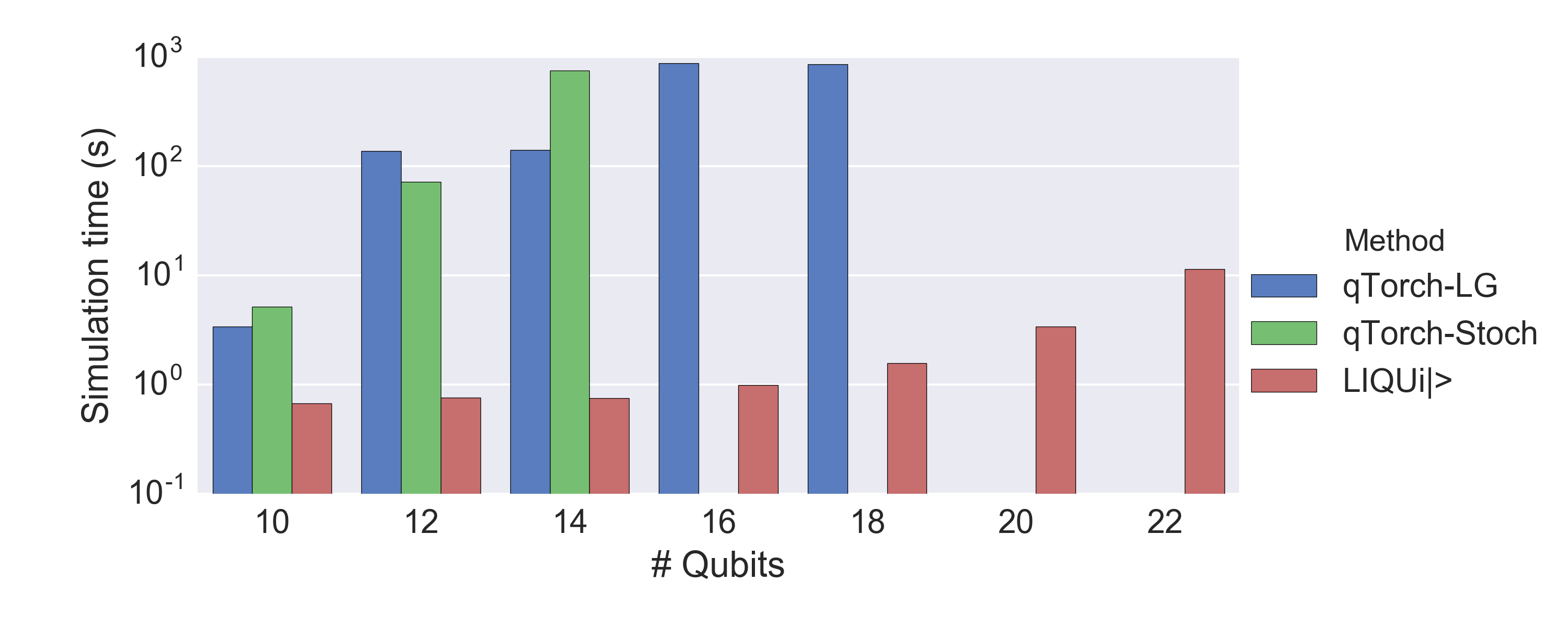}
	\caption{Time results for simulating quantum circuits of the Hubbard model. LG, Stoch, and \liq \space denote linegraph-based tensor contraction, stochastic tensor contraction, and \liq, respectively. \liq's full Hilbert simulation method is substantially faster than either tensor contraction method. Missing data points resulted from exceeding memory capacity.}
    \label{fig:hubb}
\end{figure}

\liq \space is the fastest simulation method to use for the Hubbard simulations, as shown in Fig \ref{fig:hubb}. This is because the treewidth of the circuit's graph increases substantially with the number of qubits, even for these short-depth circuits. The result is not surprising---if the algorithm were easy to simulate with a tensor network on a classical computer, then it would not have been worth proposing as a candidate for a quantum computer.

\begin{figure}[H]
	\centering
 	\includegraphics[width=\textwidth]{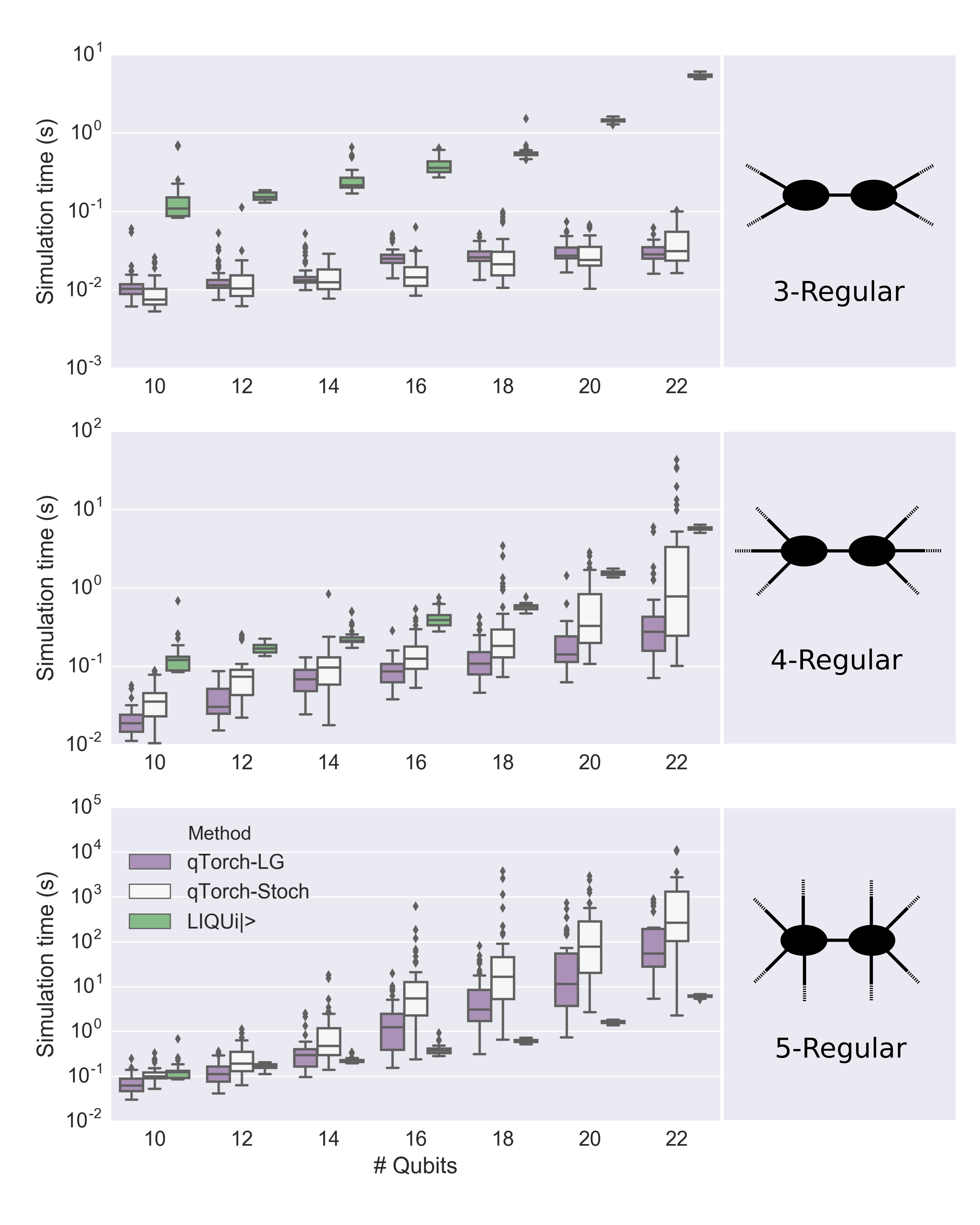}
	\caption{Simulation time plotted against number of qubits for Max-Cut/QAOA circuits. LG, Stoch, and \liq \space denote linegraph-based tensor contraction, stochastic tensor contraction, and the \liq \space software package, respectively. Tree decompositions for the LG method were determined by running the QuickBB simulation for 3000 seconds. For lower regularities, the tensor contraction methods outperform \liq, since \liq \space simulates the full Hilbert space. However, as the regularity of the Max-Cut graphs (and hence the treewidth of the quantum circuits' line graphs) increase, full Hilbert space simulation using \liq \space becomes more efficient.}
    \label{fig:t-nq-few}
\end{figure}

\begin{figure}[H]
	\centering
 	\includegraphics[width=\textwidth]{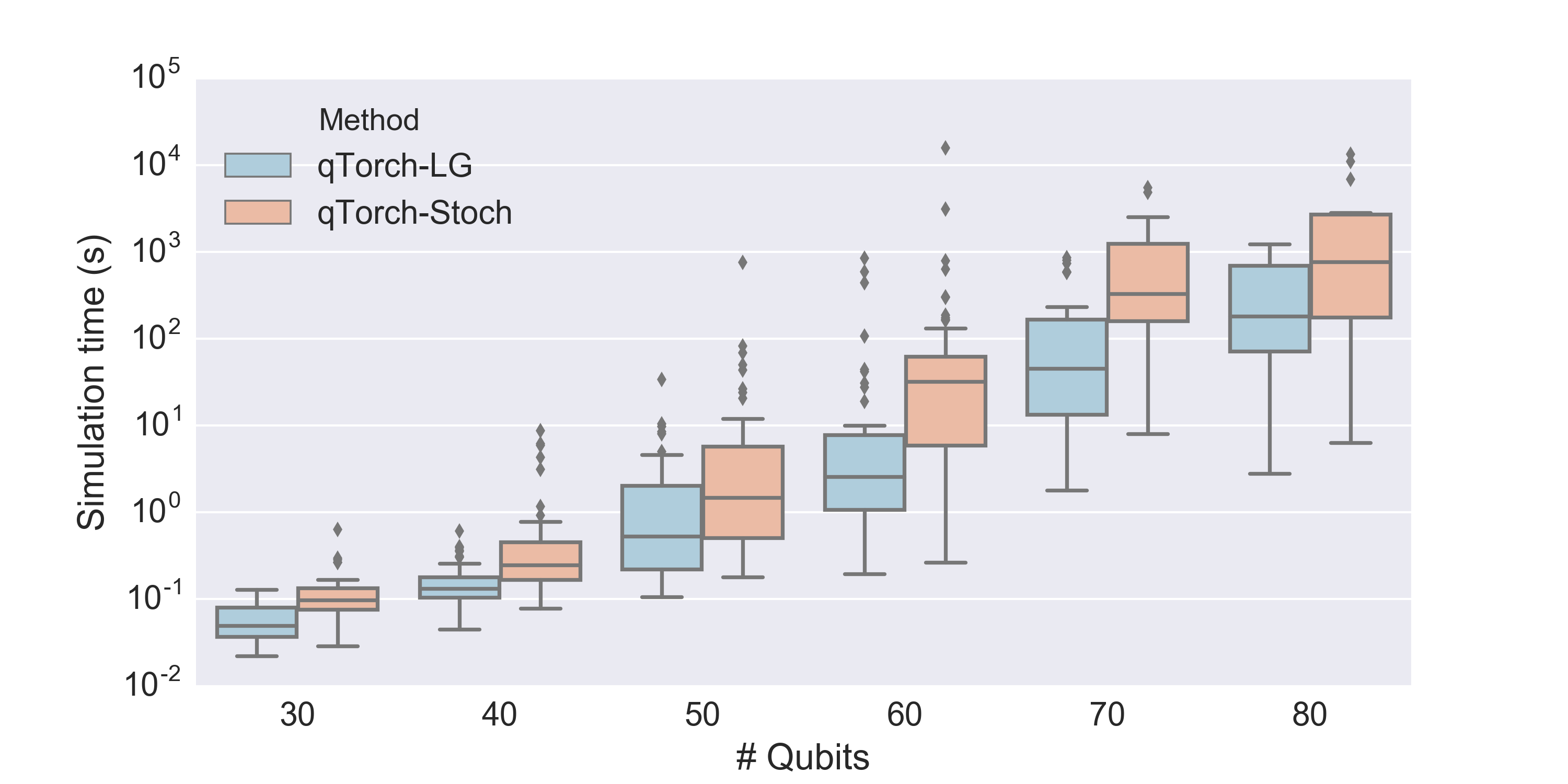}
	\caption{Simulation time plotted against number of qubits for 3-regular Max-Cut/QAOA circuits. LG and Stoch denote linegraph-based tensor contraction and stochastic tensor contraction respectively. For 3-regular Max-Cut/QAOA circuits, we were able to simulate a small subset of the 100-qubit circuits we created, not shown here.}
    \label{fig:t-nq-many}
\end{figure}

Simulation timing results for 3-, 4-, and 5-regular Max-Cut/QAOA circuits are shown using Tukey boxplots in Figs \ref{fig:t-nq-few} and \ref{fig:t-nq-many}. Stoch and LG simulation times are of similar order of magnitude for these circuits, though LG is generally faster. The exception is the 3-regular graph problems, where Stoch often appears to find a more efficient contraction than the 3000-second run of QuickBB does. We note that if the QuickBB algorithm were run for infinite time before beginning the contraction, then qTorch-LG should always (except in very simple graphs) contract the circuit faster than qTorch-Stoch. This is because, while the Stoch search is purely local (considering only individual wires), the tree decomposition approach of QuickBB implicitly considers the effects of multiple contraction steps. Note that actual search time of Stoch is negligible compared to the tensor contraction time. Note that \liq \space begins to outperform tensor contraction methods once the algorithm is run on 5-regular graphs, because the increased circuit complexity leads to larger intermediate tensors in qTorch.

Note that in principle, Hilbert space simulation can be considered a subset of TN contraction, where the state vector is simply a large tensor. Hence one might expect that there would not be a crossover point at all, \textit{i.e.} that in the worst case TN contraction would not ever be slower than Hilbert space simulation. However, because our implementation considers density matrices instead of state vectors, one would in fact expect this crossover point to exist. The largest tensor in qTorch would have $4^{N}$ entries, while the state vector has just $2^N$ entries. The various choices made in software implementations for qTorch and \liq \space would also affect the position of this crossover point.

Using a single Cori NERSC node, we were able to contract quantum circuits of 90 qubits for a very small subset of the simulated graphs, though not on enough graphs to report statistics. Full Hilbert space methods would be limited to $\sim$30 qubits on these nodes, and indeed previous simulation packages have not yet surpassed 46 qubits \cite{qhipster,haner-steiger2017}, using thousands of nodes. 

\begin{figure}[H]
	\centering
 	\includegraphics[width=0.8\textwidth]{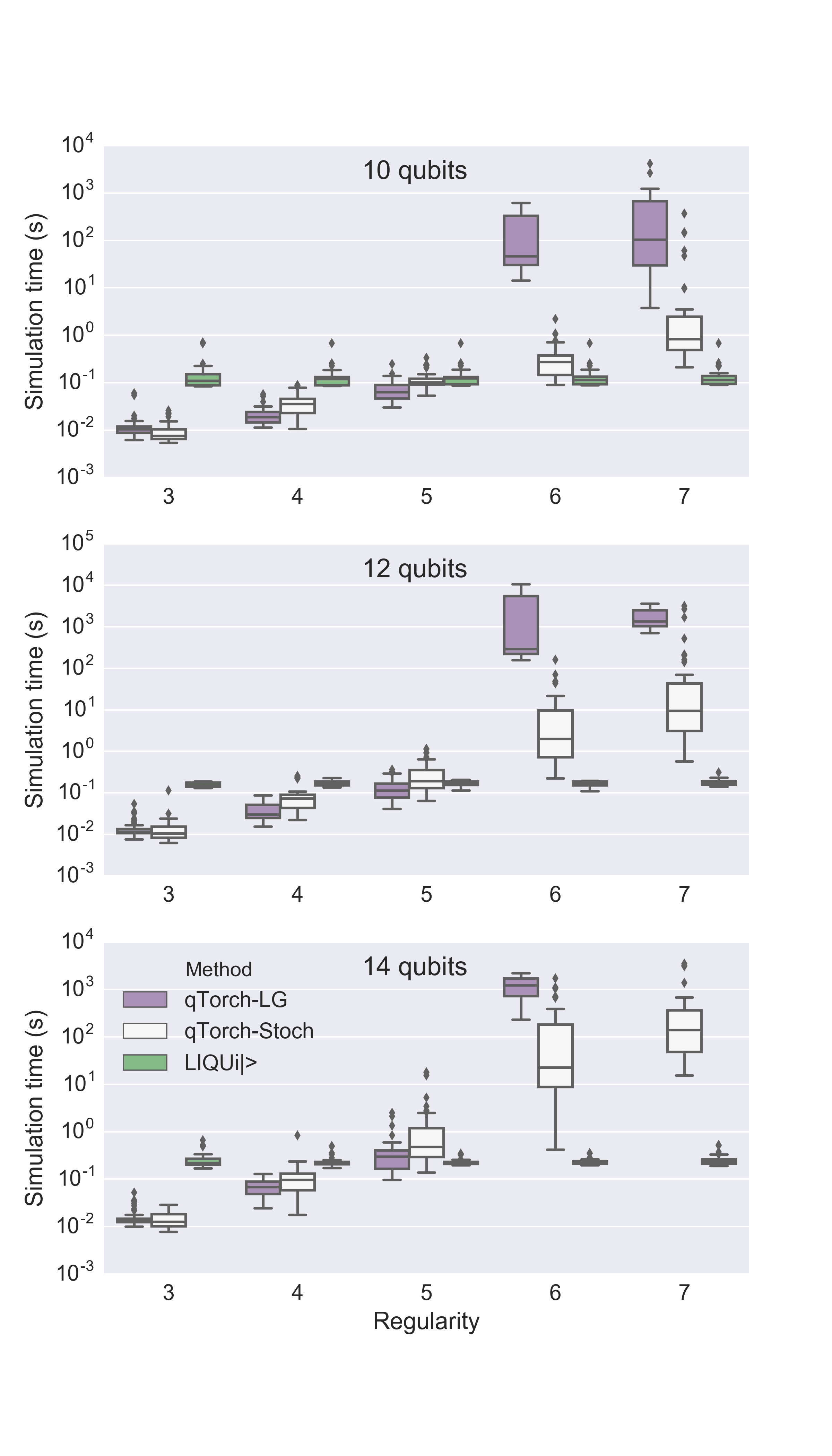}
	\caption{Simulation time plotted against the regularity of the underlying Max-Cut graph, for Max-Cut/QAOA circuits. LG, Stoch, and \liq \space denote linegraph-based tensor contraction, stochastic tensor contraction, and \liq, respectively. As regularity increases, full Hilbert space simulation (using \liq \space) becomes a more competitive simulation method. Missing data points resulted from running out of memory..}
    \label{fig:t-reg}
\end{figure}

Interesting trends appear when the simulation time is plotted against regularity of the Max-Cut problem's graph (Fig \ref{fig:t-reg}). It is notable that the LG method runs out of memory before the Stoch method does. As previously mentioned, the LG method contracts more efficiently the longer QuickBB has been run, and we chose 3000 seconds as an arbitrary QuickBB limit for all circuits. There is a trade-off between running a longer QuickBB simulation and instead immediately using the Stoch method. Even with few qubits, at higher regularities the full Hilbert space simulation (using \liq) performs better. This is expected, since as the complexity of the quantum circuit increases, higher-rank intermediate tensors appear.

\begin{figure}[H]
    \hspace{-1in}
 	\includegraphics[scale=0.43]{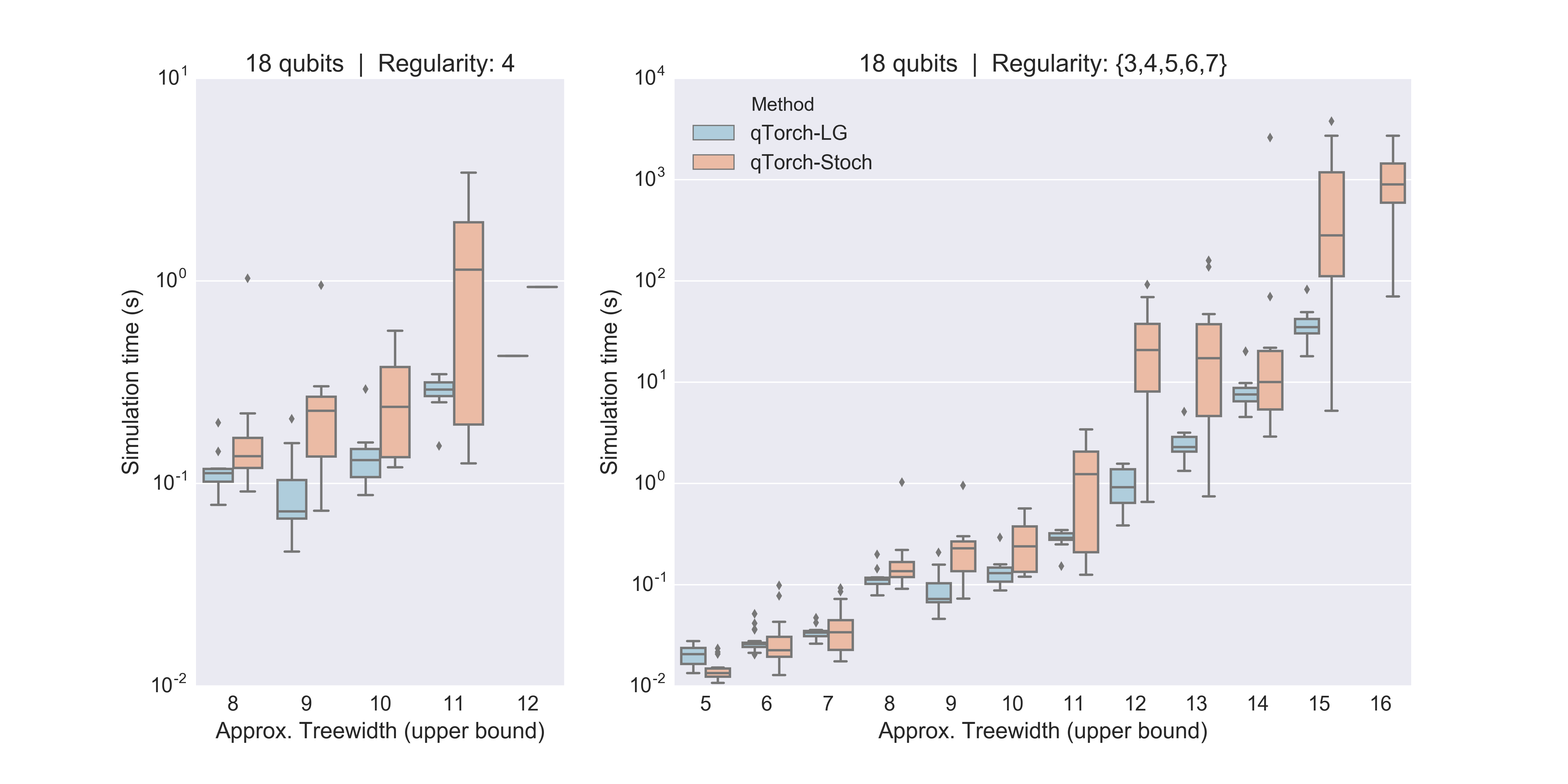}
	\caption{Simulation time plotted against approximate treewidth, for all simulated Max-Cut/QAOA quantum circuits of 18 qubits. The plot demonstrates the general trend of increased simulation time with the quantum circuit's line graphs's treewidth, despite a constant number of qubits.}
    \label{fig:t-tw}
\end{figure}

Fig \ref{fig:t-tw} shows simulation time as the estimated upper bound for the treewidth increases, for Max-Cut/QAOA circuits of 18 qubits. These include 3- through 7-regular graphs. This treewidth upper bound is simply the treewidth of the tree decomposition that defines the contraction ordering. The plot shows the expected general trend of an increase simulation time with increased treewidth, regardless of contraction scheme.

Finally, we note that we were easily able to perform simulations of 100 qubits for less complex graphs. To report one such example, we produced a random 3-regular graph with a slightly different procedure from that given in of Section \ref{ssec:main-qaoa}. Beginning with a 2-regular graph (i.e. a ring) of 100 vertices, we added edges between random pairs of vertices until all vertices were of 3 degrees. Contracting this graph's Max-Cut/QAOA circuit took $\sim$150 seconds.

\section{Conclusions}
\label{sec:concl}
We have implemented a tensor contraction code for the efficient simulation of quantum circuits. We compared a stochastic contraction scheme to one based on the line graph of the quantum circuit's graph, showing that the latter is more efficient in most circuits simulated herein. However, there is a subset of cases for which calculating an efficient approximate optimal tree decomposition of the line graph takes longer than contracting the circuit stochastically, in which case the stochastic scheme is superior. For the circuits studied in this work, our simulations suggest that the point at which qTorch is no longer faster than \liq \space occurs in QAOA/Max-Cut approximately when the Max-Cut graph has a regularity of five. In the future, qTorch may be used to estimate these points of equivalent computational cost in other classes of circuits, which may help to determine which simulation method to use in simulations.

Several immediate algorithmic improvements are possible for this software. The use of sparse tensors would reduce the number of floating point operations for some relevant circuits. Tensor decompositions (such as the singular value decomposition) along with trimming can be added as intermediate steps, as has been done in tensor network based simulations of physical systems\cite{orus13,ran17}. Additionally, more advanced parallelization methods would allow for faster calculation of a tree decomposition as well as faster contractions.

\section*{Supporting information}


\section*{Acknowledgements}
We are grateful to Edward Farhi and Aram Harrow for discussions about QAOA, to Salvatore Mandr\`a for general discussions, and to Dave Wecker for helpful advice on using \liq. N.S. acknowledges the use of resources of the National Energy Research Scientific Computing Center, a DOE Office of Science User Facility supported by the Office of Science of the U.S. Department of Energy under Contract No. DE-AC02-05CH11231. A.A.-G. and E.F. were supported by the Office of Naval Research under grant N00014-16-1-2008 (Vannevar Bush Faculty Fellowship). A. A.-G. and Y.C. were supported by NSF grant CHE-1655187.


\begin{thebibliography}{10}

\bibitem{Liquid}
Wecker D, Svore KM. LIQUi$|>$: A Software Design Architecture and
  Domain-Specific Language for Quantum Computing; 2014.
\newblock arXiv:1402.4467.

\bibitem{qhipster}
Smelyanskiy M, Sawaya NPD, Aspuru-Guzik A. {qHiPSTER: The Quantum High
  Performance Software Testing Environment}; 2016.
\newblock arXiv:1601.07195.

\bibitem{qcmpi}
Tabakin F, Julia-Diaz B.
\newblock Qcmpi: A Parallel Environment for Quantum Computing.
\newblock Computer Physics Communications. 2009;180:948.
\newblock doi:{10.1016/j.cpc.2008.11.021}.

\bibitem{qcwave}
Tabakin F, Juliá-Díaz B.
\newblock \{QCWAVE\} – A {Mathematica} quantum computer simulation update.
\newblock Computer Physics Communications. 2011;182(8):1693 -- 1707.
\newblock doi:{http://dx.doi.org/10.1016/j.cpc.2011.04.010}.

\bibitem{1012.6035}
Miszczak JA.
\newblock Models of quantum computation and quantum programming languages.
\newblock Bull Pol Acad Sci-Tech Sci. 2010;59(3):305.
\newblock doi:{10.2478/v10175-011-0039-5}.

\bibitem{qhip-jctc}
Sawaya NPD, Smelyanskiy M, McClean JR, Aspuru-Guzik A.
\newblock Error Sensitivity to Environmental Noise in Quantum Circuits for
  Chemical State Preparation.
\newblock Journal of Chemical Theory and Computation. 2016;12(7):3097--3108.
\newblock doi:{10.1021/acs.jctc.6b00220}.

\bibitem{silva2008}
Silva M, Magesan E, Kribs DW, Emerson J.
\newblock Scalable protocol for identification of correctable codes.
\newblock Phys Rev A. 2008;78:012347.
\newblock doi:{10.1103/PhysRevA.78.012347}.

\bibitem{geller2013}
Geller MR, Zhou Z.
\newblock Efficient error models for fault-tolerant architectures and the Pauli
  twirling approximation.
\newblock Phys Rev A. 2013;88:012314.
\newblock doi:{10.1103/PhysRevA.88.012314}.

\bibitem{tomita2014}
Tomita Y, Svore KM.
\newblock Low-distance surface codes under realistic quantum noise.
\newblock Phys Rev A. 2014;90:062320.
\newblock doi:{10.1103/PhysRevA.90.062320}.

\bibitem{boixo18a}
Boixo S, Isakov SV, Smelyanskiy VN, Babbush R, Ding N, Jiang Z, et~al.
\newblock Characterizing quantum supremacy in near-term devices.
\newblock Nature Physics. 2018;14(6):595--600.
\newblock doi:{10.1038/s41567-018-0124-x}.

\bibitem{farhi16}
{Farhi} E, {Harrow} AW. Quantum Supremacy through the Quantum Approximate
  Optimization Algorithm; 2016.
\newblock arXiv:1602.07674.

\bibitem{boixo18b}
Boixo S, Isakov SV, Smelyanskiy VN, Neven H. Simulation of low-depth quantum
  circuits as complex undirected graphical models; 2017.
\newblock arXiv:1712.05384.

\bibitem{chen18a}
Chen J, Zhang F, Huang C, Newman M, Shi Y. Classical Simulation of
  Intermediate-Size Quantum Circuits; 2018.
\newblock arXiv:1805.01450.

\bibitem{childs17}
Childs AM, Maslov D, Nam Y, Ross NJ, Su Y. Toward the first quantum simulation
  with quantum speedup; 2017.
\newblock arXiv:1711.10980.

\bibitem{bouland18}
Bouland A, Fefferman B, Nirkhe C, Vazirani U. Quantum Supremacy and the
  Complexity of Random Circuit Sampling; 2018.
\newblock arXiv:1803.04402.

\bibitem{pednault17}
Pednault E, Gunnels JA, Nannicini G, Horesh L, Magerlein T, Solomonik E,
  et~al.. Breaking the 49-Qubit Barrier in the Simulation of Quantum Circuits;
  2017.
\newblock arXiv:1710.05867.

\bibitem{chen18b}
Chen ZY, Zhou Q, Xue C, Yang X, Guo GC, Guo GP. 64-Qubit Quantum Circuit
  Simulation; 2018.
\newblock arXiv:1802.06952.

\bibitem{projq}
{Steiger} DS, {H{\"a}ner} T, {Troyer} M. {ProjectQ: An Open Source Software
  Framework for Quantum Computing}; 2016.
\newblock arXiv:1612.08091.

\bibitem{haner-steiger2017}
H\"aner T, Steiger DS. 0.5 Petabyte Simulation of a 45-Qubit Quantum Circuit;
  2017.
\newblock arXiv:1704.01127.

\bibitem{Gottesman_1998}
Gottesman D. The {H}eisenberg Representation of Quantum Computers; 1998.
\newblock arXiv:quant-ph/9807006.

\bibitem{quant-ph/0406196}
Aaronson S, Gottesman D.
\newblock Improved Simulation of Stabilizer Circuits.
\newblock Physical Review A. 2004;70:052328.
\newblock doi:{10.1103/PhysRevA.70.052328}.

\bibitem{Valiant:2001:QCS:380752.380785}
Valiant LG.
\newblock Quantum Computers That Can Be Simulated Classically in Polynomial
  Time.
\newblock In: Proceedings of the Thirty-third Annual ACM Symposium on Theory of
  Computing. STOC '01. New York, NY, USA; 2001. p. 114--123.
\newblock Available from: \url{http://doi.acm.org/10.1145/380752.380785}.

\bibitem{quant-ph/0108010}
Terhal BM, DiVincenzo DP.
\newblock Classical simulation of noninteracting-fermion quantum circuits.
\newblock Physical Review A. 2002;65:032325.
\newblock doi:{10.1103/PhysRevA.65.032325}.

\bibitem{2016PhRvL.116y0501B}
{Bravyi} S, {Gosset} D.
\newblock Improved Classical Simulation of Quantum Circuits Dominated by
  Clifford Gates.
\newblock Physical Review Letters. 2016;116(25):250501.
\newblock doi:{10.1103/PhysRevLett.116.250501}.

\bibitem{6657072}
García HJ, Markov IL.
\newblock Quipu: High-performance simulation of quantum circuits using
  stabilizer frames.
\newblock In: 2013 IEEE 31st International Conference on Computer Design
  (ICCD); 2013. p. 404--410.

\bibitem{garcia17}
García HJ, Markov IL.
\newblock Simulation of Quantum Circuits via Stabilizer Frames.
\newblock IEEE Transactions on Computers. 2017;64(8).

\bibitem{gould-PI-2006}
Rudiak-Gould B. The sum-over-histories formulation of quantum computing; 2006.
\newblock arXiv:quant-ph/0607151.

\bibitem{quant-ph/0403114}
Viamontes GF, Markov IL, Hayes JP.
\newblock Graph-based simulation of quantum computation in the density matrix
  representation.
\newblock Quantum Information \& Computation. 2005;5(2):113.
\newblock doi:{10.1117/12.542767}.

\bibitem{quant-ph/0511069}
Markov IL, Shi Y.
\newblock Simulating quantum computation by contracting tensor networks.
\newblock SIAM Journal on Computing. 2005;38(3):963.
\newblock doi:{10.1137/050644756}.

\bibitem{mikeike11}
Nielsen MA, Chuang IL.
\newblock Quantum Computation and Quantum Information: 10th Anniversary
  Edition.
\newblock 10th ed. New York, NY, USA: Cambridge University Press; 2011.

\bibitem{McCaskey_2016}
McCaskey AJ. {Tensor Network Quantum Virtual Machine (TNQVM)}; 2016.
\newblock Available from:
  \url{http://www.osti.gov/scitech/servlets/purl/1340180}.

\bibitem{orus13}
Or{\'{u}}s R.
\newblock A practical introduction to tensor networks: Matrix product states
  and projected entangled pair states.
\newblock Annals of Physics. 2014;349:117--158.
\newblock doi:{10.1016/j.aop.2014.06.013}.

\bibitem{orus14}
Or{\'{u}}s R.
\newblock {Advances on tensor network theory: Symmetries, fermions,
  entanglement, and holography}.
\newblock European Physical Journal B. 2014;87(11).
\newblock doi:{10.1140/epjb/e2014-50502-9}.

\bibitem{ran17}
Ran SJ, Tirrito E, Peng C, Chen X, Su G, Lewenstein M. Review of Tensor Network
  Contraction Approaches; 2017.
\newblock arXiv:1708.09213.

\bibitem{Robertson1991153}
Robertson N, Seymour PD.
\newblock {Graph minors. X. Obstructions to tree-decomposition}.
\newblock Journal of Combinatorial Theory, Series B. 1991;52(2):153 -- 190.
\newblock doi:{http://dx.doi.org/10.1016/0095-8956(91)90061-N}.

\bibitem{Bodlaender2010259}
Bodlaender HL, Koster AMCA.
\newblock {Treewidth computations I. Upper bounds}.
\newblock Information and Computation. 2010;208(3):259 -- 275.
\newblock doi:{http://dx.doi.org/10.1016/j.ic.2009.03.008}.

\bibitem{BODLAENDER20111103}
Bodlaender HL, Koster AMCA.
\newblock {Treewidth computations II. Lower bounds}.
\newblock Information and Computation. 2011;209(7):1103 -- 1119.
\newblock doi:{http://dx.doi.org/10.1016/j.ic.2011.04.003}.

\bibitem{qbb}
Gogate V, Dechter R.
\newblock A Complete Anytime Algorithm for Treewidth.
\newblock In: Proceedings of the 20th Conference on Uncertainty in Artificial
  Intelligence. UAI '04. Arlington, Virginia, United States: AUAI Press; 2004.
  p. 201--208.
\newblock Available from:
  \url{http://dl.acm.org/citation.cfm?id=1036843.1036868}.

\bibitem{Arnborg:1987:CFE:37170.37183}
Arnborg S, Corneil DG, Proskurowski A.
\newblock Complexity of Finding Embeddings in a K-tree.
\newblock SIAM J Algebraic Discrete Methods. 1987;8(2):277--284.
\newblock doi:{10.1137/0608024}.

\bibitem{amir10}
Amir E.
\newblock Approximation Algorithms for Treewidth.
\newblock Algorithmica. 2010;56(4):448--479.
\newblock doi:{10.1007/s00453-008-9180-4}.

\bibitem{pfeifer14}
Pfeifer RNC, Haegeman J, Verstraete F.
\newblock Faster identification of optimal contraction sequences for tensor
  networks.
\newblock Physical Review E. 2014;90(3).
\newblock doi:{10.1103/physreve.90.033315}.

\bibitem{Peruzzo.NC.5.4213.2014}
Peruzzo A, McClean J, Shadbolt P, Yung MH, Zhou XQ, Love PJ, et~al.
\newblock A variational eigenvalue solver on a photonic quantum processor.
\newblock Nat Commun. 2014;5:4213.

\bibitem{weckervqe15}
Wecker D, Hastings MB, Troyer M.
\newblock Progress towards practical quantum variational algorithms.
\newblock Phys Rev A. 2015;92:042303.
\newblock doi:{10.1103/PhysRevA.92.042303}.

\bibitem{Yung.SR.4.3589.2014}
Yung MH, Casanova J, Mezzacapo A, McClean J, Lamata L, Aspuru-Guzik A, et~al.
\newblock {From transistor to trapped-ion computers for quantum chemistry}.
\newblock Sci Rep. 2014;4:3589.
\newblock doi:{10.1038/srep03589}.

\bibitem{Mcclean.NJP.18.023023.2016}
McClean JR, Romero J, Babbush R, Aspuru-Guzik A.
\newblock The theory of variational hybrid quantum-classical algorithms.
\newblock New J Phys. 2016;18(2):023023.

\bibitem{kraus83}
Kraus K, Böhm A, Dollard JD, Wootters WH, editors.
\newblock States, Effects, and Operations Fundamental Notions of Quantum
  Theory.
\newblock Springer Berlin Heidelberg; 1983.
\newblock Available from: \url{https://doi.org/10.1007%2F3-540-12732-1}.

\bibitem{aharonov06}
Aharonov D, Kitaev A, Preskill J.
\newblock Fault-Tolerant Quantum Computation with Long-Range Correlated Noise.
\newblock Phys Rev Lett. 2006;96:050504.
\newblock doi:{10.1103/PhysRevLett.96.050504}.

\bibitem{preskill13}
Preskill J.
\newblock Sufficient Condition on Noise Correlations for Scalable Quantum
  Computing.
\newblock Quantum Info Comput. 2013;13(3-4):181--194.

\bibitem{farhi14a}
{Farhi} E, {Goldstone} J, {Gutmann} S. {A Quantum Approximate Optimization
  Algorithm}; 2014.
\newblock arXiv:1411.4028.

\bibitem{farhi14b}
{Farhi} E, {Goldstone} J, {Gutmann} S. A Quantum Approximate Optimization
  Algorithm Applied to a Bounded Occurrence Constraint Problem; 2014.
\newblock arXiv:1412.6062.

\bibitem{wecker16}
{Wecker} D, {Hastings} MB, {Troyer} M.
\newblock {Training a quantum optimizer}.
\newblock Phys Rev A. 2016;94(2):022309.
\newblock doi:{10.1103/PhysRevA.94.022309}.

\bibitem{lin16}
{Yen-Yu Lin} C, {Zhu} Y. Performance of QAOA on Typical Instances of Constraint
  Satisfaction Problems with Bounded Degree; 2016.
\newblock arXiv:1601.01744.

\bibitem{gg17}
{Giacomo Guerreschi} G, {Smelyanskiy} M. {Practical optimization for hybrid
  quantum-classical algorithms}; 2017.
\newblock arXiv:1701.01450.

\bibitem{nlopt}
Johnson SG. {The NLopt nonlinear-optimization package};.
\newblock \url{http://ab-initio.mit.edu/nlopt}.

\bibitem{mcclean2016hybrid}
McClean JR, Kimchi-Schwartz ME, Carter J, de~Jong WA.
\newblock Hybrid quantum-classical hierarchy for mitigation of decoherence and
  determination of excited states.
\newblock Phys Rev A. 2017;95:042308.
\newblock doi:{10.1103/PhysRevA.95.042308}.

\bibitem{Omalley.PRX.6.031007.2016}
O'Malley PJJ, Babbush R, Kivlichan ID, Romero J, McClean JR, Barends R, et~al.
\newblock Scalable Quantum Simulation of Molecular Energies.
\newblock Phys Rev X. 2016;6:031007.
\newblock doi:{10.1103/PhysRevX.6.031007}.

\bibitem{openfermion}
McClean JR, Kivlichan ID, Sung KJ, Steiger DS, Cao Y, Dai C, et~al..
  {OpenFermion}: The Electronic Structure Package for Quantum Computers; 2017.
\newblock arXiv:1710.07629.

\bibitem{Whitfield.MP.109.735.2011}
Whitfield JD, Biamonte J, Aspuru-Guzik A.
\newblock {Simulation of electronic structure Hamiltonians using quantum
  computers}.
\newblock Mol Phys. 2011;109(5):735--750.
\newblock doi:{10.1080/00268976.2011.552441}.

\bibitem{docker2014}
Merkel D.
\newblock Docker: Lightweight Linux Containers for Consistent Development and
  Deployment.
\newblock Linux J. 2014;2014(239).

\end{thebibliography}
\end{document}